\def\year{2019}\relax
\documentclass[letterpaper]{article} 
\usepackage{bm}
\usepackage{svg}
\usepackage{cite}
\usepackage{amsmath,amssymb,amsfonts}
\usepackage{algorithmic}
\usepackage{graphicx}
\usepackage{textcomp}
\usepackage{xcolor}
\usepackage{placeins}
\usepackage{booktabs}
\usepackage{siunitx}
\usepackage{threeparttable}
\usepackage{pdfpages}
\usepackage[version=4]{mhchem}
\usepackage{footnote}

\usepackage{etoolbox}
\usepackage{aaai19}  
\usepackage{times}  
\usepackage{helvet}  
\usepackage{courier}  
\usepackage{url}  
\usepackage{graphicx}  
\begin{document}
\author{
Chengqiang Lu$^\dag${\normalfont,\;}
Qi Liu$^\dag$\thanks{Contact author.}{\normalfont,\;}
Chao Wang$^\dag${\normalfont,\;}
Zhenya Huang$^\dag${\normalfont,\;}
Peize Lin$^\ddag${\normalfont,\;}
Lixin He$^\ddag$ \\
	$^\dag$Anhui Province Key Lab. of Big Data Analysis and Application, University of S\&T of China\\$^\ddag$Key Laboratory of Quantum Information, University of S\&T of China
	\\ \{qiliuql, helx\}@ustc.edu.cn, \{lunar, wdyx2012, huangzhy, linpz\}@mail.ustc.edu.cn}
	

\title{Molecular Property Prediction: A Multilevel Quantum Interactions\\ Modeling Perspective}
\maketitle

\begin{abstract}
Predicting molecular properties (e.g., atomization energy) is an essential issue in quantum chemistry, which could speed up much research progress, such as drug designing and substance discovery. Traditional studies based on density functional theory (DFT) in physics are proved to be time-consuming for predicting large number of molecules. Recently, the machine learning methods, which consider much rule-based information, have also shown potentials for this issue. However, the complex inherent quantum interactions of molecules are still largely underexplored by existing solutions. In this paper, we propose a generalizable and transferable Multilevel Graph Convolutional neural Network (MGCN) for molecular property prediction. Specifically, we represent each molecule as a graph to preserve its internal structure. Moreover, the well-designed hierarchical graph neural network directly extracts features from the conformation and spatial information followed by the multilevel interactions. As a consequence, the multilevel overall representations can be utilized to make the prediction. Extensive experiments on both datasets of equilibrium and off-equilibrium molecules demonstrate the effectiveness of our model. Furthermore, the detailed results also prove that MGCN is generalizable and transferable for the prediction.
\end{abstract}

\section{Introduction}
Predicting molecular properties, such as atomization energy, is one of the fundamental issues in quantum chemical science.
Indeed, it has attracted much attention in relevant fields of physics, chemistry and computer science, since it speeds up the societal and technological progress in the application of discovering substances with desired characteristics, such as drug design with specific target and new material manufacture~\cite{becke2007quantum,oglic2017active}.

In the literature, density functional theory (DFT) plays an important role in physics for molecular property prediction. It holds a common statement that the quantum interactions between particles (e.g., atom) create the correlation and entanglement of molecules which are closely related to their inherent properties~\cite{thouless2014quantum}. Along this line, many quantum mechanical methods based on DFT have been developed to model the quantum interactions of molecules for the prediction~\cite{hohenberg1964inhomogeneous,kohn1965self}. However, DFTs are computationally costly since they usually use specific functions to determine the interactions of particles, which proves to be extraordinarily time consuming. For example, experimental results indicated that it took nearly an hour to predict the properties of merely one molecule with 20 atoms~\cite{Gilmer2017NeuralMP}. Obviously, it is unacceptable to make prediction on large number of molecules in chemical compound space. Therefore, it is necessary to find more effective solutions. 

\begin{figure}[tb]
	\centerline{\includegraphics[width=.95\columnwidth]{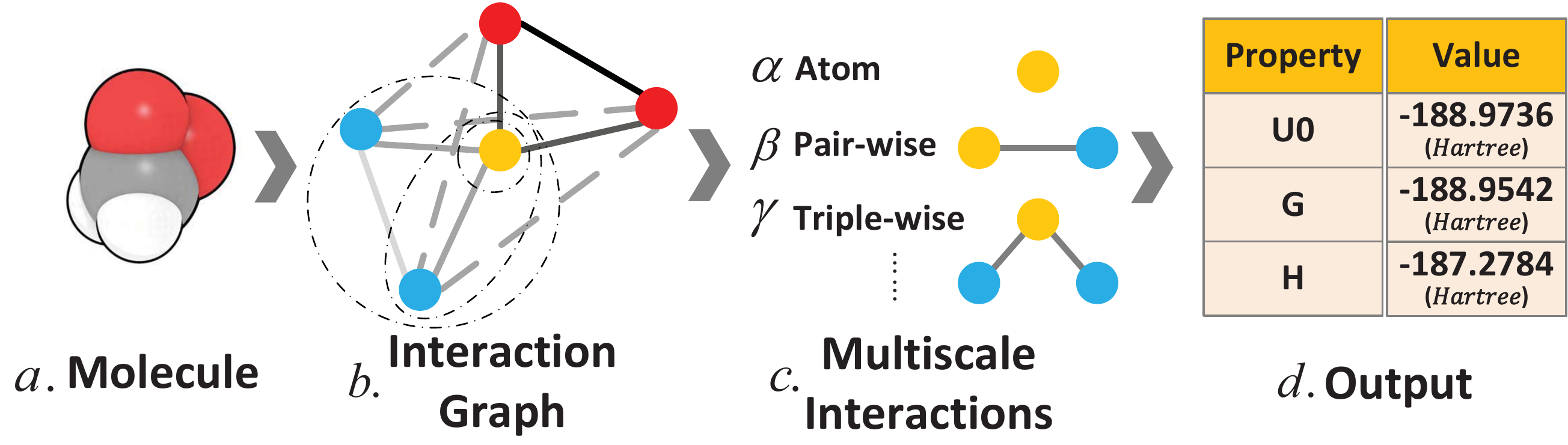} }
	\caption{Illustration of the process of a molecule (\ce{CH2O2}) via our method.}
	\label{demo}
\end{figure}

Recently, inspired by the remarkable success of machine learning in many tasks including computer vision, natural language processing, natural and social science ~\cite{Karpathy2014LargeScaleVC,He2016DeepRL,huang2017question,zhu2018xiaoice,liu2018finding}, researchers have shown the potentials of these data-driven techniques for molecular property prediction~\cite{faber2017prediction,schutt2017schnet}. Generally, these studies mainly rely on rule-based feature engineering (e.g., bag of atom bonds) or treat molecules as grid-like structures (e.g., 2D images or text). However, few of them directly take the inherent quantum interactions of molecules into consideration, causing severe information loss, which makes the molecular property prediction problem pretty much open.

Unfortunately, there are many technical and domain challenges along this line. First, there are highly complex quantum interactions, such as distracted attraction, exchange repulsion and electrostatic interaction in molecules, especially in the large molecules~\cite{kollman1985theory}. It is hard to model them with analytical methods. Second, compared with traditional tasks including computer vision, the amount of labeled molecule data is significantly limited, which requires a generalizable approach for the prediction. Last but not least, in practice, we are often provided with labeled data of small and medium molecules except large molecules since the calculation of them are expensive. Thus, it is necessary to notice this unbalancedness to propose a transferable solution for property prediction of large molecules using the model trained on smaller ones.

To address these challenges, in this paper, we propose a well-designed Multilevel Graph Convolutional Neural Network (MGCN) for predicting properties of molecules by directly incorporating their quantum interactions. Figure ~\ref{demo} demonstrates the process of our approach. Specifically, we first represent each molecule as an interaction graph, which could preserve its internal structure without information loss. Then we propose a hierarchical graph convolutional neural network to model the multilevel quantum interactions based on the graph-like molecular structures. Here, we follow  the DFT theory that the quantum interactions could be transformed at different levels, i.e., atom-wise refers to the inherent influence of each atom (e.g., oxygen), atom-pair refers to the interaction between two atoms, atom-triple means the correlation among three atoms, and so on. Thus, our proposed graph network incorporates hierarchical layers of point-wise, pair-wise, triple-wise, etc to extract representations of the multilevel interactions, respectively. Finally, the overall interaction representation from all levels could be utilized to make the property (e.g., atomization energy) prediction. We conduct extensive experiments on both datasets of equilibrium and off-equilibrium molecules, where the experimental results shows the effectiveness of our proposed approach. Moreover, as MGCN could naturally pass the interaction information of molecules level by level, which also proves the superior ability of generalizability and transferability.

\section{Related Work}
Generally, the related work of our research could be classified into the following three categories.

\textbf{Density Functional Theory.}
Molecular property prediction problem has been studied for a long time in physics, chemistry and material science~\cite{wang2011application}. In the literature, density functional theory (DFT) is the most popular method, which plays a vital role in making the prediction, and could date back to 1960s~\cite{hohenberg1964inhomogeneous,kohn1965self,lawless2002information}. Generally, it states that the quantum interactions between particles (e.g., atoms) create the correlation and entanglement of molecules which are closely related to their inherent properties~\cite{thouless2014quantum}. Following this theory, many DFT based methods, such as B3LYP, were proposed, which mapped the quantum interactions of molecules onto every single particles, for predicting the properties~\cite{yanai2004new}. However, the complexity of DFT could be approximated as $\mathcal{O}(N^3)$, where $N$ denotes the number of particles. Therefore, it is time-consuming in the experiments and unacceptable for the prediction when facing large number of molecules~\cite{Gilmer2017NeuralMP}.


\textbf{Traditional Machine Learning Methods.}
To find more efficient solutions for molecular property prediction, researchers have attempted to leverage various machine learning models, such as kernel ridge regression, random forest and Elastic Net \cite{faber2017prediction,zou2005regularization,mcdonagh2017machine}. Generally, they rely on rule-based hand crafted features using the domain knowledge of physics and chemistry, including bag of bonds, coulomb matrix, and histogram of distances, angles and dihedral angles~\cite{huang2016communication,hansen2015machine,montavon2012learning}. Although some superior experimental results have been achieved, these traditional machine learning methods take manual feature engineering, which requires much domain expertise. Thus, they are often restricted in practice.

\textbf{Deep Neural Networks.}
Compared to traditional machine learning models, deep neural networks hold a superiority of automatic feature learning, which have achieved great success in many applications, such as speech recognition~\cite{zhu2016co}, computer vision~\cite{lecun1995convolutional} and natural language processing~\cite{collobert2008unified}. With this ability, researchers have noticed the potentials of these deep methods for molecular property prediction. Along this line, convolutional neural network based models were proposed, where they represented each molecule as grid-like structures, such as image~\cite{goh2017chemnet}, string~\cite{GmezBombarelli2018AutomaticCD}, and sphere~\cite{boomsma2017spherical}. For example, Goh et al.~\shortcite{goh2017chemnet} converted molecular diagrams into 2D RGB images and proposed the ChemNet for the prediction. However, this grid-like transformation usually caused information loss of the molecules which lied in non-Euclidean space, where the internal spatial and distance information of atoms were not fully considered \cite{bronstein2017geometric}. Therefore, some works operated the molecule as a atom graph and developed graph convolutional neural networks for the property prediction~\cite{schutt2017quantum,Gilmer2017NeuralMP}. For instance, Schutt et al.~\shortcite{schutt2017quantum} proposed the deep tensor neural network that captured the representation of each atom node in molecules. Shang et al.~\shortcite{shang2018edge} further introduced attention mechanism for characterizing the edge information to improve the prediction.

Our work improves the previous studies as follows. First, we propose the multilevel graph network to directly model the multilevel quantum interactions of molecules from hierarchical perspectives (i.e., atom-wise, pair-wise, triple-wise, etc), which developed the graph modeling for molecular property prediction. Second, our work could pass the interaction information level-by-level, which benefits more practical scenarios, i.e., generalizability of limited data and transferability of unbalanced data.

\section{Multilevel Graph Convolutional Network}
In this section, we first formally introduce the molecular property prediction problem. Then we describe our Multilevel Graph Convolutional Network in detail.

\subsection{Problem Statement}
Given a molecule, it is natural to represent it as graphs without the loss of information, where vertices represent atoms and edges represent chemical bonds. Thus, a molecule is denoted by $\mathcal{G} \langle \mathcal{V}, \mathcal{E} \rangle$, and in the setting of molecular structure, $\mathcal{V}$  is a set of atoms with $|\mathcal{V}|=N$. We regard the graph as a complete undirected graph following the assumption that every atom has the interactions with others so that the set of edges satisfies that $|\mathcal{E}| = N(N-1)/2 $. Here each $\mathcal{E}$ contains two kinds of information, namely edge type and spatial information, respectively. Our target is to construct a regressor to predict the properties of molecules. Formally, we can define the problem as:  
 \begin{equation}
    g(f(\mathcal{G})) = y,  
\end{equation}   
where $y$ is the target property to predict and the middle function $f:\mathcal{G}\rightarrow \mathbb{R}^{N\scriptsize{\times}  D}$ is used to learn representations of atoms. Then $g$ converts the obtained features to final result.   

The multilevel interactions widely exist in the graph structures. In the field of molecule, physical experts design the different symmetry functions to describe the atomic environment by considering the interactions at varied levels \cite{behler2014representing}. Inspired by this idea, we model quantum interactions in molecules by representing the interactions between two, three, and more atoms level by level to demystify the complexity of molecular interactions. In the next subsection, we will introduce our Multilevel Graph Convolutional Network (MGCN) in detail.

\subsection{Network Architecture}  
  
\begin{figure*}[tb]
\centerline{\includegraphics[width=17.3cm]{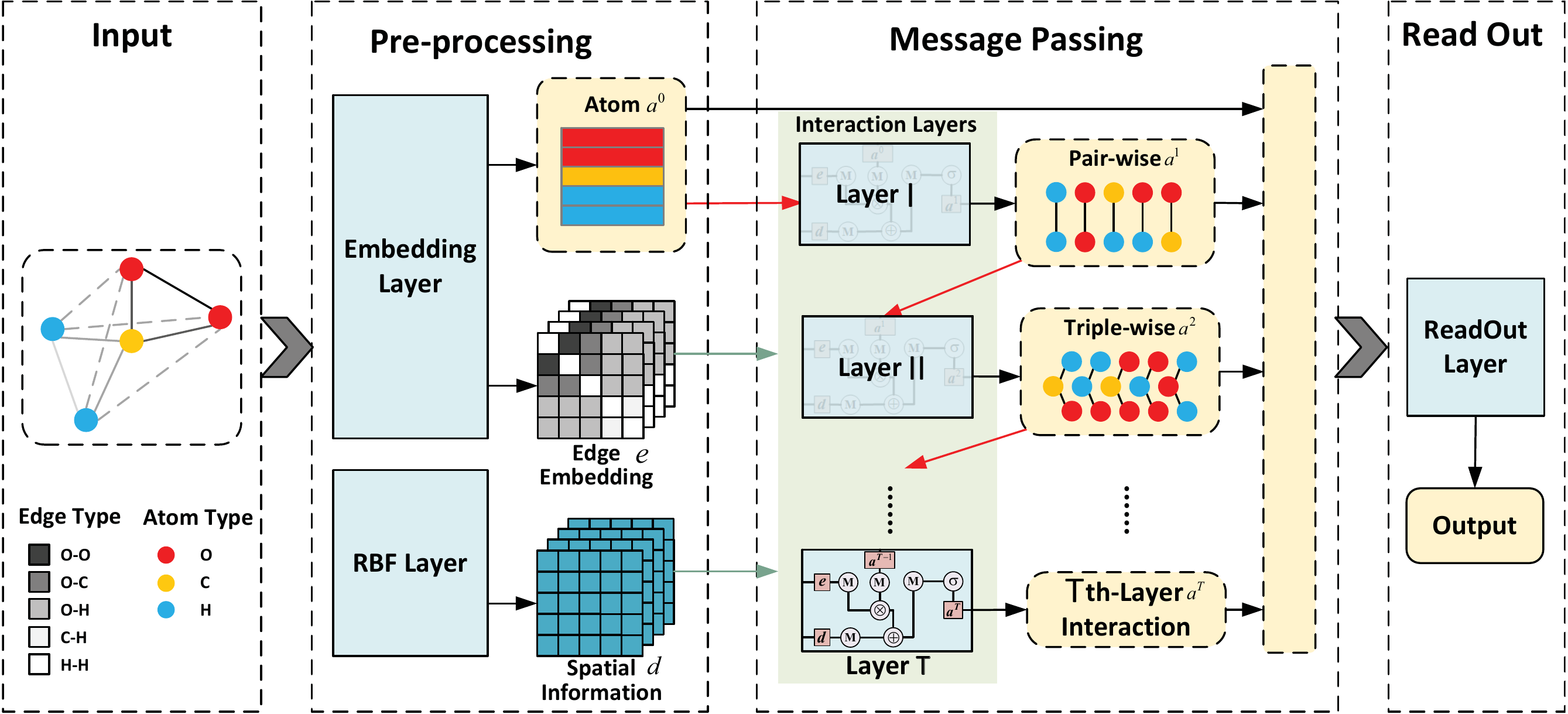}}
\caption{The architecture of the entire MGCN.} 
\label{network}
\end{figure*}

\textbf{Overview.}\ The entire architecture could be split into three parts in a high-level discussion except for thse input. The initial input is a graph which consists of a list of atoms and a Euclidean distance matrix of the molecules. The pre-processing part includes embedding layer and Radial Basis Function (RBF) layer. The embedding layer generates atom and edge embeddings while Radial Basis Function (RBF) layer converts the distance matrix to a distance tensor. The next part of MGCN are several interaction layers that aim to learn different node representations in different levels. The last phase is the readout layer that outputs the final result. 

\textbf{Embedding Layer.}\ Atoms and bonds are the basic elements in a molecule. Thus, to model interactions with as less information loss as possible, we present an embedding layer to directly embed vertices and edges of a graph into vectors.
Each atom in a molecule is represented as a vector $\bm{a}^0\scriptsize{\in} \mathbb{R}^D$ initially. Therefore, the vertices in the entire molecular are denoted as a matrix $ A^0 \scriptsize{\in} \mathbb{R}^{N\scriptsize{\times}  D}$ and $\bm{a}^0_i$ indicates the atom embedding of $i$-th atom in a molecule. The atoms that have the same number of protons in their atomic nuclei share the same initial representation which is called the atom embedding here. Taking \ce{CH2O2} as an example, there are five atoms and different kinds of atoms are labeled with different colors in the input part of Figure \ref{network}. After the process of embedding layer, we get a $5 \scriptsize{\times}  D$ matrix and the rows that are related to the atoms of the same type share the same value. The atom embeddings of all chemical elements are generated randomly before training. The initialization of pair-wise embeddings $\bm{e}\scriptsize{\in} \mathbb{R} $ is similar to atom embeddings (see the preprocess part of Figure \ref{network}). Thus we get $E\scriptsize{\in} \mathbb{R}^{N \scriptsize{\times}  N \scriptsize{\times}  D}$, and the edges connecting the same set of atoms have the same initial edge embedding. Specifically, $\bm{e}_{ij}$ indicates the edge embedding of the bond between $i$-th atom and $j$-th atom. The representations generated by the embedding layer are only related to the inherent property of isolated atoms and bonds. The interaction terms are modeled in later subnetwork.

\textbf{Radial Basis Function Layer.}\ The spatial information influences the degree of interactions between nodes and we use the RBF Layer to convert these information to robust distance tensors for further utilization. First of all, we reform the raw coordinates of atoms to distance matrix to remove the disturbance of selection of coordinate frame. Secondly, Radial Basis Functions are applied to convert the distance matrix to a distance tensor.

RBF is a widespread kernel method which originally was invented to generate function interpolation \cite{broomhead1988radial}. Its variant was proved to be advantageous to create fingerprint-like descriptor of molecules \cite{Li2018DeeperII}. Here we use RBF to spread the 2D inter-atomic distance matrix to a 3D representation. Given a set of $K$ central points $\{ \mu_1,\dots,\mu_K \}$, the single data point $x$, namely one pair-wise distance in the molecule, will be processed as:   

 \begin{equation} 
	\mathit{RBF}(x) = \mathop{\frown}\limits_{i=1}^{K}\mathrm{h}(\|x-\mu_i\|). 
\end{equation}

\noindent Here the notation $\frown$ means concatenation, and we take Euclidean distance as the norm. As for radial basis function $\mathrm{h}$, we take Gaussian $\exp(-\beta\|x-\mu_i\|^2)$ following the suggestion in \cite{schutt2017schnet} to avoid the long plateau on the initial phases of the training procedure. $k$ central points are picked evenly in the range from the shortest to the longest edge among the entire dataset. Therefore all distances in the dataset will be covered. 

Through the non-linear transformation, the representations of distances between nodes become more robust. Furthermore, more additional interpretation is introduced by radial basis function layer than simple multi-layer perceptron. After the RBF layer, we create the pair-wise distance tensors $D \scriptsize{\in} \mathbb{R}^{N \scriptsize{\times}  N \scriptsize{\times}  K}$, and $d_{ij}$ denotes the distance tensor between $i$-th atom and $j$-th atom.

\textbf{Interaction Layer.}\ To model the multilevel molecular structure with all the conformation and spatial information embedded through previous layers, we construct the interaction layer which is a crucial component of our model. Considering that the quantum interactions in molecules could be transformed at different levels (i.e., atom-wise, atom-pair, atom-triple, etc), our interaction layer is designed by the hierarchical architecture level by level. Specifically, in the $l$-th interaction layer, we define the edge representation $e_{ij}^{l+1}$ and atom representation $a^{l+1}_i$ as:
 \begin{equation}\bm{e}^{l+1}_{ij} = \mathrm{h}_{e}(\bm{a}_i^l, \bm{a}_j^l, \bm{e}^l_{ij}), \end{equation}   
  \begin{equation}\bm{a}^{l+1}_i = \sum_{\substack{j=1, j\neq i}}^N \mathrm{h}_{v}(\bm{a}^l_j, \bm{e}^l_{ij}, \bm{d}_{ij}), \end{equation}   
where $h_e$ is used to update edge representation and $h_v$ is the function that collects the message from the neighbours of the $i$-th atom to generate $\bm{a}_i^{l+1}$. 

With this hierarchical modeling, MGCN could effectively preserve the structure of each molecules and describe its quantum interactions. Specifically, in the first layer, $\bm{a}^0_i$ denotes the atom embedding that show the inherent properties of certain chemical elements. As the forward inference steps, $\bm{a}^1_{i}$ involves the first-order neighbour node and spatial information with the message passed by $\bm{a}^0$, $e$ and $d$. In a similar way, $\bm{a}^2_{i}$ represents the triple-wise interactions, $\bm{a}^3_{i}$ indicates the interactions between four nodes and so on. As shown in Figure.\ref{network}, after each interaction layer, we obtain the representations of atoms that reflect the higher-order interactions thanks to decomposition of molecule. 

The update function $\mathrm{h}_e$ is calculated as:  

  \begin{equation} 
\mathrm{h}_e = \eta \bm{e}^l_{ij} \oplus (1 - \eta)W^{ue} \bm{a}^{l}_i\odot\bm{a}^{l}_j. \end{equation} where $\eta$ is the hyper-parameter that controls the influence of former pair-wise information (default value is 0.8). Here $\odot$ and $\oplus$ denote the element-wise dot and plus respectively. In this way, The edge embedding is corrected by the related atomic representations of former interaction layer.

The function $\mathrm{h}_v$ applies the message passing operation to create the atom representation at a higher order. The distance tensor $d_{ij}$ here controls the magnitude of impact in each pair of atoms and the edge embedding $e_{ij}$ provides the extra bond information. Thus, it combines the information of nodes, edges, and space, more formally:  
  \begin{equation}
\mathrm{h}_v = \sigma(W^{uv} ( \mathrm{M}^{fa}(\bm{a}^l_j) \odot \mathrm{M}^{fd}(\bm{d}_{ij}) \oplus \mathrm{M}^{fe}(\bm{e}_{ij})), 
\end{equation} 
where $\sigma$ is a tanh activation function, $W^{uv}$ is a weight matrix. The notation $\mathrm{M}(\bm{x})$ refers to a dense layer that $\mathrm{M}(x)=W\bm{x} +\bm{b}$ with the input $\bm{x}$ for simplicity. $\mathrm{M}^{fa}$,$\mathrm{M}^{fd}$,$\mathrm{M}^{fe}$ are dense layers here.

\textbf{Readout Layer.}\ After the interaction layers, we get atom representations at different levels. In the last phase, we construct a readout layer to make the final prediction utilizing these features more clearly.   

First of all, we aggregate the various atom representations to obtain the final vertex feature map as following:  

  \begin{equation}
\bm{a}_i = \mathop{\frown}\limits_{k=0}^{T}\bm{a}_i^k, 
\end{equation}   

\noindent where $T$ indicates the number of interaction layers and $\mathop{\frown}$ means concatenation.    

Secondly, we need to predict the property of molecule with the multilevel representations of each atoms. Fortunately, the molecular properties satisfy additivity and locality. For example, to predict the energetic property, we can model potential energy surfaces as follows \cite{behler2014representing,cubuk2017representations}:  

  \begin{equation}
E=\sum_{i}^{N}\sum_{j}^{N}E_{ij}, 
\end{equation}  

\noindent where $E$ is the total energy and $E_{ij}$ indicates the part of energy related to the bond between $i$-th and $j$-th atom ($i\neq j$). Besides, $E_{ii}$ could be regarded as the partial energy that mapped to $i$-th atom. Along this line, we can process the representations separately and then sum them up:   
  \begin{equation}
\hat{y} =\sum_{i=1}^N W^{r^a_{2}}\sigma(\mathrm{M}^{r^a_{1}}(\bm{a}_i)) +  \mathop{\sum\limits_{i=1}^N\sum\limits_{j=1}^N}\limits_{i\neq j} W^{r^e_{2}}\sigma(\mathrm{M}^{r^e_{1}}(\bm{e}_{ij})), 
\end{equation}     
\noindent where $\sigma$ is the activation function, more specifically, the softplus function. The former term refers to the contribution of quantum interactions that mapped to each atom. Additionally, the latter term denotes the edge-related contribution that can not be mapped to single particle. Since the atom-related interactions account for vast majority of molecular interactions, the latter term is nuanced. Therefore, when the amount of data is small, we tend to ignore the latter term.

 To train this model, we use the Root-Mean-Square Error (RMSE) as our loss function: 
   \begin{equation}
     \ell(\hat{y},y)=\sqrt{|\hat{y}-y|^2},
 \end{equation}   
 where $\hat{y}$ denotes the predictive value and $y$ is the true value.

\subsection{Discussion on MGCN} 

\textbf{Generalizability.}\   
In the field of chemistry, the set of all possible molecules in unexplored regions is called chemical space. One of the famous chemical space project \cite{Ruddigkeit2012EnumerationO1} collected 166.4 billion molecules while merely 134k samples of them were labeled \cite{ramakrishnan2015many}. Therefore, the generalization ability of enabling accurate prediction with the limited dataset is indeed essential in our task.

In the design of our model, we decide to use the distance tensors $D$ as the form of spatial information instead of coordinates of atom. Accordingly, MGCN enforces rotation and translation invariance. Henceforth, the representation learned by MGCN is more general and would not be confused with the same molecule in different orientations.  

Moreover, we perform element-wise operations in interaction layers (equation(5) and (6)) to generate representations and process the representations of each atom  respectively. Under those circumstances, the prediction made by our model is irrelevant to sequence of atoms. The index invariance enhances the generalization ability of MGCN.

Additionally, we utilize some normalization techniques such as dropout to prevent overfitting, which also benefits generalizability of our model.

In brief, our model is generalizable which is particular important for molecular property prediction where the amount of training data is limited.

\textbf{Transferability.}\ Since the expensive computational cost is a critical bottleneck which limits capabilities to calculate the properties of large molecules, most open data are small and medium molecules and the amount of large molecules is small. Therefore, the ability of transferring the knowledge learned from small molecules to larges ones could help us deal with the data-hungry of big molecules.  

The atom/edge embeddings generated by embedding layer are only in regard to the type of atoms and edges and irrelevant to the specific molecular structure and 
spatial information. As a result, the chemical-domain knowledge learned in the embeddings is universal in the molecular system no matter small or large molecules. Then, in our mulitilevel phase, we use the embeddings to generate the representation in deeper level, e.g., pair-wise and triple-wise. Although small and large molecules are different in the distribution of atoms and bonds, their interactions in different levels are similar. Consequently, with the general embeddings and similar interaction mechanism, MGCN could infer the higher-level representations to predict property and maintain a certain accuracy. Therefore, our model that trained on the small molecules could obtain competitive performance in the prediction of larger molecules.

Rather than applying the model trained on small molecules to big molecules directly, another way to transfer the knowledge is using pre-trained embeddings. To train a model in large molecules, we could initialize this model with the atom and edge embeddings of another model that was trained on small molecules. The pre-trained embeddings could speed up the convergence and improve the accuracy, because the domain knowledge in embeddings learned from small molecules is still meaning suitable to big molecules.

Along this line, this model is capable of transferring the knowledge of small molecules to large molecules and tackle the structural shortage of data.

\textbf{Time Complexity.}\ 
The time complexity of our MGCN model is $\mathcal{O}(N^2_a)$ since the calculation of $h_e$ and $h_v$ (equation (5) and (6)) are independent of molecular size. Here $N_a$ indicates the number of atoms in a molecule.

\section{Experiments}

We conduct experiments to demonstrate the effectiveness of MGCN from various aspects: 1) predictive performance; 2) the effectiveness of multilevel structure; 3) the validation of generalizability; 4) the verification of transferability; 5) the influence of varied number of interaction layers.

\subsection{Datasets}

\textbf{QM9.}\ The QM9\footnote{http://www.quantum-machine.org/datasets/\#qm9} dataset \cite{ramakrishnan2014quantum} is perhaps the most well-known benchmark dataset which contains  134k equilibrium molecules with their 13 different properties. All of the relaxed geometries and properties for all the 134k molecules are calculated by DFT. The DFT error is the empirical inaccuracy estimation of DFT based approaches \cite{faber2017prediction}. The QM9 dataset also provides the chemical accuracy which is generally accepted by the chemistry community as a relatively ideal accuracy. 

\begin{table*}[tbp]
\centering
\footnotesize

\caption{Predictive accuracy of different models in QM9}
\begin{threeparttable}
 \resizebox{2\columnwidth}{!}{

\begin{tabular}{lrrrrrrrrrrrrr}
    \toprule
           Properties & \multicolumn{1}{l}{$U_0$} & \multicolumn{1}{l}{$U$} & \multicolumn{1}{l}{$G$} & \multicolumn{1}{l}{$H$} & \multicolumn{1}{l}{$C_v$} & \multicolumn{1}{l}{ $\varepsilon_{\text{HOMO}}$} & \multicolumn{1}{l}{$\varepsilon_{\text{LUMO}}$} & \multicolumn{1}{l}{$\Delta \varepsilon$} & \multicolumn{1}{l}{$\omega_1$} & \multicolumn{1}{l}{ZPVE} & \multicolumn{1}{l}{$\langle R^2 \rangle $} & \multicolumn{1}{l}{$\mu$} & \multicolumn{1}{l}{$\alpha$} \\
           Unit & \multicolumn{1}{l}{eV} & \multicolumn{1}{l}{eV} & \multicolumn{1}{l}{eV} & \multicolumn{1}{l}{eV} & \multicolumn{1}{l}{cal/molK} & \multicolumn{1}{l}{eV} & \multicolumn{1}{l}{eV} & \multicolumn{1}{l}{eV} & \multicolumn{1}{l}{$\text{cm}^{-1}$} & \multicolumn{1}{l}{eV} & \multicolumn{1}{l}{$\text{Bohr}^2$} & \multicolumn{1}{l}{Debye} & \multicolumn{1}{l}{$\text{Bohr}^3$} \\
    \midrule

    DFT Error & 0.1    & 0.1    & 0.1    & 0.1    & 0.34   & -      & -      & -      & 28     & 0.0097 & -      & 0.1    & 0.4 \\
    Chemical Acc. & 0.043  & 0.043  & 0.043  & 0.043  & 0.05   & 0.043  & 0.043  & 0.043  & 10     & 0.00122 & 1.2    & 0.1    & 0.1 \\
    \midrule
    \midrule
    RF+BAML & 0.2000  & -      & -      & -      & 0.451  & 0.1070  & 0.1180  & 0.1410  & 2.71   & 0.01320  & 51.10  & 0.434  & 0.638  \\
    KRR+BOB & 0.0667  & -      & -      & -      & 0.092  & 0.0948  & 0.1220  & 0.1480  & 13.20  & 0.00364  & 0.98   & 0.423  & 0.298  \\
    KRR+HDAD & 0.0251  & -      & -      & -      & 0.044  & 0.0662  & 0.0842  & 0.1070  & 23.10  & 0.00191  & 1.62   & 0.334  & 0.175  \\
    GG     & 0.0421  & -      & -      & -      & 0.084  & 0.0567  & 0.0628  & 0.0877  & 6.22   & 0.00431  & 6.30   & 0.247  & 0.161  \\
    enn-s2s & 0.0194  & 0.0194 & 0.0168 & 0.0189 & 0.040  & 0.0426  & 0.0374 & 0.0688 & 1.90   & 0.00152  & 0.18   & \textbf{0.030} & 0.092  \\
    DTNN   & 0.0364  & 0.0377 & 0.0385 & 0.0357 & 0.089  & 0.0982  & 0.1053  & 0.1502  & 4.23   & 0.00312  & 0.30   & 0.257  & 0.131 \\
    SchNet & 0.0134  & 0.0189 & 0.0196 & 0.0182 & 0.067  & 0.0507  & \textbf{0.0372}  & 0.0795  & 3.83   & 0.00172  & 0.27   & 0.071  & 0.073  \\
    MGCN   & \textbf{0.0129} & \textbf{0.0144} & \textbf{0.0146} & \textbf{0.0162} & \textbf{0.038} & \textbf{0.0421} & 0.0574  & \textbf{0.0642}&\textbf{1.67} & \textbf{0.00112} & \textbf{0.11} & 0.056  & \textbf{0.030}  \\
    \bottomrule
    
    \end{tabular}}

\end{threeparttable}

\label{qm9}%
\end{table*}%

\begin{table}[htbp]
  \centering
  \caption{Predictive accuracy of different models in ANI-1}
    \begin{tabular}{lrrrr}
    \toprule
    Methods & \multicolumn{1}{l}{DTNN} & \multicolumn{1}{l}{SchNet} & \multicolumn{1}{l}{MGCN}  \\
    \midrule
    
    MAE    & 0.113  & 0.108   & \textbf{0.078} \\
    \bottomrule
    \end{tabular}%
  \label{ani}%
\end{table}%

\noindent \textbf{ANI-1.}\ The ANI-1\footnote{https://www.nature.com/articles/sdata2017193} dataset provides access to the total energies of 20 million off-equilibrium molecules which is 100 times larger than QM9.

\subsection{Experimental Setup}  
 
We use mini-batch stochastic gradient descent (mini-batch SGD) with the Adam optimizer \cite{Kingma2014Adam} to train our MGCN. The batch size is set to 64 and the initial learning rate is 1$e^{-5}$. For all 13 properties of QM9, we pick 110k out of 130k molecules randomly as our training set that accounts for about 84.7\% of the entire dataset. With the rest of the data, we choose half of them as the validation set and the other half as the testing set. As for the much larger ANI-1, we randomly choose 90\% samples for training, 5\% samples for validation and 5\% for testing. We select Mean Absolute Error (MAE) as our evaluation metrics for the convenience of comparison with baselines \cite{faber2017prediction}.

\subsection{Baselines}  

We compare our model with the 7 baseline methods that could be categorized into two groups.   

The first group consist of 3 traditional ML models using hand-engineered features derived from the molecular literature \cite{faber2017prediction,huang2016communication,hansen2015machine}. These ML models include Random Forest (RF) and Kernel Ridge Regression (KRR). The hand-craft features include Bag of Bonds (BOB), Bond-Angle Machine Learning (BAML) and "Projected Histograms" (HDAD). We imply the combination of X regressor and Y representation with the notation X+Y. Thus, these three baselines are denoted by RF+BAML, KRR+BOB, KRR+HDAD. These models achieve the best performance in the prediction of one or more properties among all 30 combinations of regressors and features \cite{faber2017prediction}. 
 
The second group contain 4 deep neural networks. They are gated graph network (GG, \citeauthor{kearnes2016molecular} \citeyear{kearnes2016molecular}), edge neural network with set-to-set (enn-s2s, \citeauthor{Gilmer2017NeuralMP} \citeyear{Gilmer2017NeuralMP}), deep tensor neural network (DTNN, \citeauthor{schutt2017schnet} \citeyear{schutt2017quantum}) and SchNet \cite{schutt2017schnet}. These models are proved to be competitive in the molecular property prediction. Noting that DTNN and SchNet only provide their experimental results in the prediction of property $U_0$ for the molecules in QM9, thus we complete the rest of the experiments. Besides, all of other numerical results of the baselines are extracted from their works directly.

\subsection{Experimental Results}
\textbf{Predictive performance.}\ We compare our model with the baseline models mentioned above in two datasets. In Table \ref{qm9}, we provide the MAE of baselines and our approach as well as DFT error and chemical accuracy for all 13 properties. Table \ref{ani} shows the performance comparison in ANI-1.

As illustrated in Table \ref{qm9}, MGCN gets the best performance in 11 out of 13 properties, and 11 of them exceed the chemical accuracy. Our model is able to improve the performance upon state-of-the-art. Another observation is that the deep neural networks (GG, enn-s2s, DTNN, SchNet and MGCN) outperform the models that using hand-craft features comprehensively. 
In the experiment in ANI-1, we choose the state-of-the-art models (DTNN and SchNet) as comparison. As shown in Table \ref{ani}, the accuracies in ANI-1 is lower than in QM9, and there are possible two reasons. First, the force in equilibrium molecules of QM9 is negligible, while in off-equilibrium molecules of ANI-1, this factor increases the complexity of quantum interactions. Second, the 100 times larger size of ANI-1 than QM9 makes it more difficult to ﬁt. Even though, our model still achieves satisfactory accuracy and outperform other methods.

In brief, our model attains the best performance benefiting from the mulitilevel interaction modeling, and the results prove that our model could handle both equilibrium and off-equilibrium molecules and is capable to fit the large dataset. Considering that most previous work have no experiment in the ANI-1 dataset, we chose QM9 as our default dataset in the rest of our experiments.

\begin{figure}[tb]
	
    \centerline{\includegraphics[width=1.014\columnwidth]{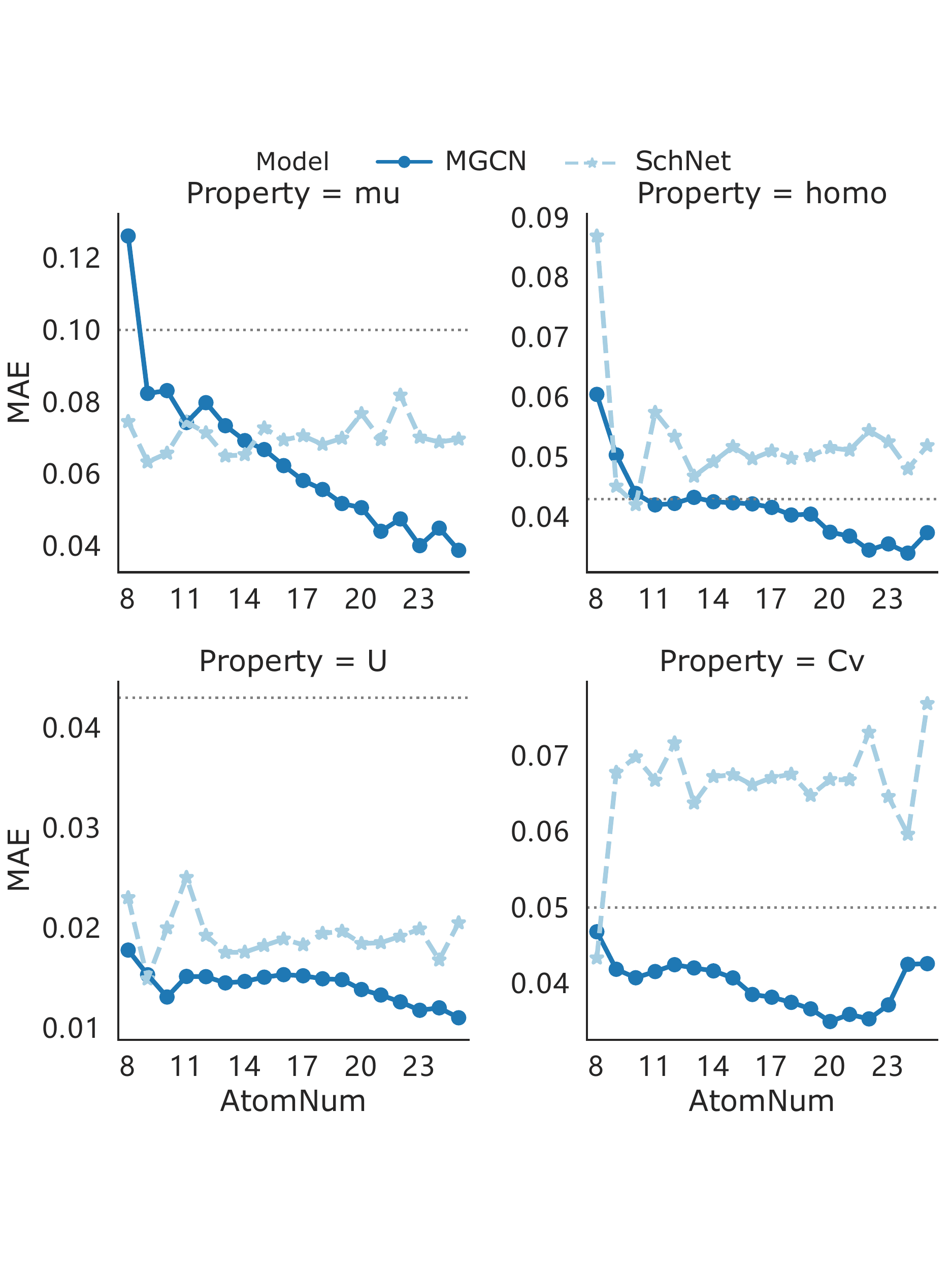}}

    \caption{MAE of prediction in different size molecules.}
    \label{maeatom}
\end{figure}

 \textbf{Effectiveness of multilevel interactions.}\ In the molecular system, with the increase of the number of atoms in a molecule, the complexity of quantum interactions will grow exponentially. In consequence, it is much harder to model the interactions of molecules if the size of them is larger.

 Figure \ref{maeatom} shows the MAE of predictions as the function of the number of atoms. We select the state-of-the-art work SchNet as a comparison for better illustration. Four representative and prevalent properties ($\mu$, $\varepsilon_{\text{HOMO}}$, $U$, $Cv$) are picked in this experiment. The dotted horizontal line in each subplot is the chemical accuracy of each property. Figure \ref{maeatom} shows that MGCN assesses more accurate and stable performance than SchNet. Furthermore, as the number of atoms increases, the advantage of MGCN becomes more apparent due to the multilevel modeling. Our model simplifies the interactions by dividing them into different levels and represent them respectively using the mulitilevel structure and decomposition of molecular quantum interactions. Along this line, our model performs better comparatively when the number of atoms increases. In addition, the significant fluctuations appearing in the front and end of the curves derive from the lack of molecules that contain less than 10 atoms or more than 24 atoms in dataset.  
 
 To investigate this further, we construct a control model that blends all levels of interactions in a single embedding rather than construct representation level by level. Taking the prediction of $U$ property for example, the result of this model is not as well as our MGCN with an MAE of 0.03683 on average. It implies the molecular representations modeled by multilevel interaction layers are more robust, which validates the effectiveness of our multilevel modeling.

\begin{table}[tbp]
\centering
\caption{Performance comparison in varied size training set}
\begin{tabular}{rrrrr}
    \toprule

    \multicolumn{1}{l}{N} & \multicolumn{1}{l}{SchNet} & \multicolumn{1}{l}{DTNN} & \multicolumn{1}{l}{enn-s2s} & \multicolumn{1}{l}{MGCN} \\
    \midrule
    \midrule
    50,000 & 0.0256 & 0.0408 &  0.0249& \textbf{0.0229} \\
    100,000 & 0.0147 & 0.0364 & - & \textbf{0.0142} \\
    110,462 & 0.0134 & -      & 0.0194 & \textbf{0.0129} \\
    \bottomrule

\end{tabular}%

\label{size}%
\end{table}%

\begin{figure}[tpb]
    \centerline{\includegraphics[width=1.014\columnwidth]{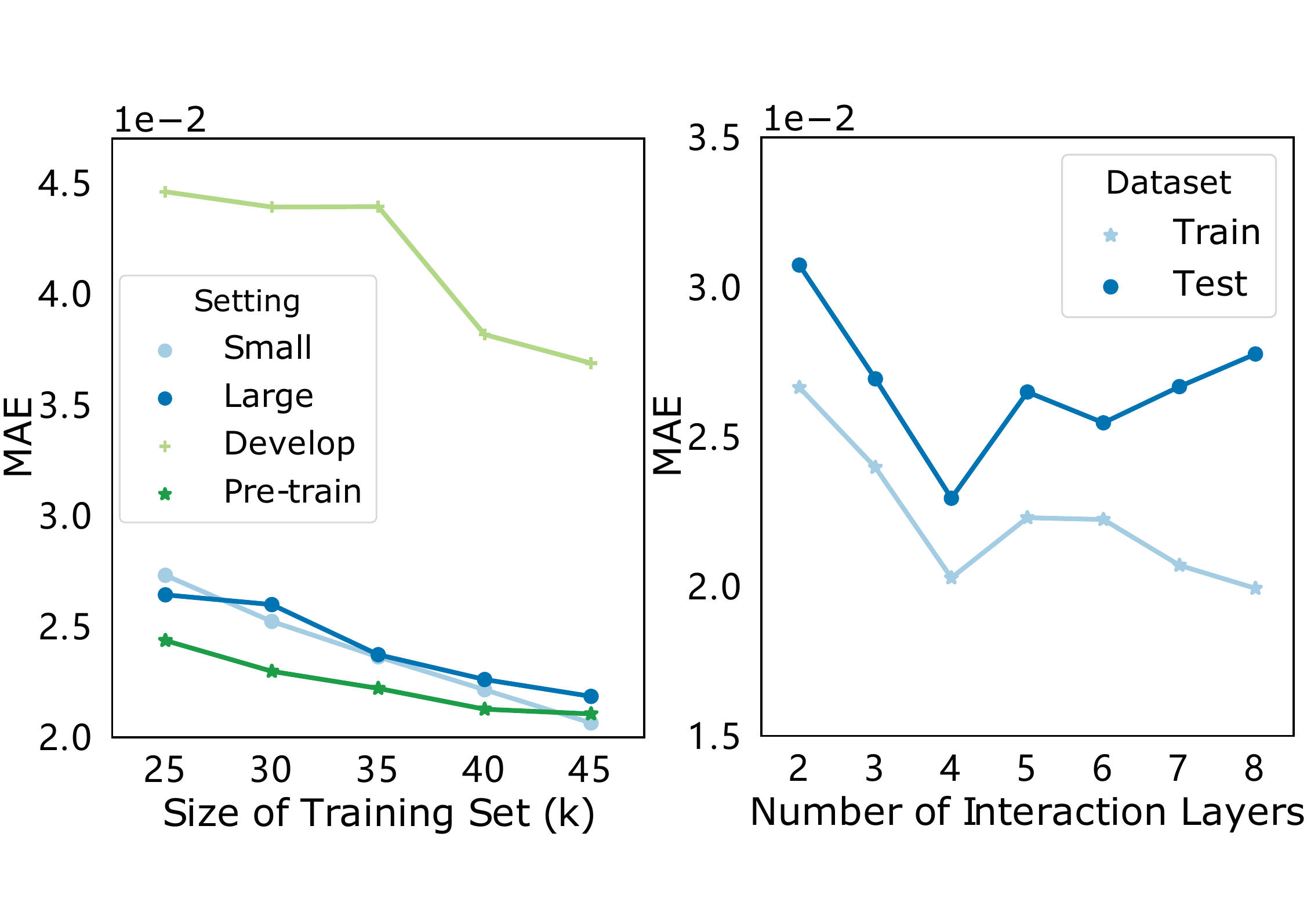}}
    \caption{$a$(left). Performance comparison in the training set with different size. $b$(right). Predictive performance of models with different number of interaction layers.}
    \label{interaction}
\end{figure}

\textbf{Generalizability.}\ The potential molecules in chemical space is numerous extra, but the amount of labeled data is quite small \cite{schutt2017quantum}. Due to the limitation of the magnitude of existing datasets, the ideal model should be able to perform well even trained with a small amount of data. Thus, the generalizability is another essential aspect to evaluate these models.  

We train our model in three training sets with the different size that consist of 50k, 100k and 110k samples respectively and test them in the same test set that contains 10k molecules. In addition to the three baselines mentioned above (DTNN, SchNet, enn-s2s), the ensemble model of five enn-s2s models is also listed in the comparison. In Table \ref{size}, the MGCN gets the lowest MAE in three training sets. Regarding that the readout phase of MGCN and SchNet are similar, the representation of molecular learned by MGCN is more generalizable when the accessible data is smaller thanks to the modeling of multilevel interactions.   

\textbf{Transferability.}\ As mentioned before, the data in existing datasets are unbalanced. For instance, relative large molecules that contain more than 20 atoms account for merely 20.7\% of total amount in QM9 and only occupy 8.5\% in ANI-1. Therefore, the transferability is quite important to an approach.  

We conduct experiments to validate the transferability of MGCN. Specifically, we sample 50k small and 50k large molecules from QM9 respectively. As shown in Figure \ref{interaction}.$a$, there are four experimental settings. The first two models are trained and tested on the small and large molecules respectively (noted by "Small" and "Large"). The latter two approaches are different ways to transfer the knowledge. The "Develop" denotes the develop model which is trained on the small molecules and applied to the large ones directly. The last model (labeled with "Pre-train") utilizes the embeddings learned on the small molecules as initialization, and then refine itself during the training on the large molecules. 

Figure \ref{interaction}.$a$ illustrates the performance of MGCN in four settings. In the first place, The MAE of small molecules is lower relatively due to the higher complexity of large molecules. Secondly, we observe that as we feed in more small data, the MAE of develop model keeps decreasing. The performance is fairly decent because we did not feed any large molecules to this model. The numerical results show that our model is capable to learn knowledge from small data and then transfer them to larger molecules. Thirdly, the pre-trained model outperforms all of other models when the size of training set is small with the universal domain knowledge learned before. This technique helps address the structural shortage of data.

\textbf{Influence of interaction layers.}\ In DFT, physicists usually use 4 to 5 different empirical symmetry functions for molecular property prediction. Each symmetry could be mapped to the interaction layer in each level. Figure \ref{interaction}.$b$ shows the relationship between the number of interaction layers and the MAE of property $U_0$. We randomly pick 50k molecules as our training set and test on remaining data. As Figure \ref{interaction}.$b$ illustrates, too many or few interaction layers could cause higher MAE. The network with less than 4 interaction layers does not have enough capacity to learn the representations of molecules and using the deeper model that contain more than 5 interaction layers will widen the generalization gap. The empirical results indicate that four is the best number of interaction layers which conform to the number of symmetry functions mentioned previously. 

\textbf{Summary.}\ Through the experiments, MGCN shows the superiority of incorporating multilevel modeling of molecular interactions. Moreover, the experimental results prove that our model is generalizable and transferable. Besides, in theory, the time complexity of MGCN is $\scriptsize{\mathcal{O}(N^2)}$ compared with $\scriptsize{\mathcal{O}(N^3)}$ of DFT. Experimentally, with the same setting (a single core of a Xeon E5-2660), our model spends $\scriptsize{2.4\scriptsize{\times}10^{-2}}$ second predicting the property of one molecule, which is nearly $\scriptsize{1.5\scriptsize{\times}10^{5}}$ times faster than DFT.

\section{Conclusion}

In this paper, we introduced a Multilevel Graph Convolutional Network (MGCN) for molecular property prediction. The well-designed model utilized the multilevel structure in molecular system to learn the representations of the quantum interactions level by level, and then made prediction with overall interaction representation. The experimental results on two prevalent datasets demonstrated the competency of our approach. Furthermore, our model was proved to be generalizable and transferable.  

We believe future research should concentrate efforts on enhancing the generalization of the atom representation because the predictive accuracy is quite high in small samples and it is tough to obtain the dataset of sufficient large molecules. 

\section{ Acknowledgments}
This research was supported by grants from the National Natural Science Foundation of China (Grants No. 61672483, 11774327), and the Science Foundation of Ministry of Education of China \& China Mobile (No. MCM20170507). Qi Liu gratefully acknowledges the support of the Young Elite Scientist Sponsorship Program of CAST and the Youth Innovation Promotion Association of CAS (No. 2014299).

\bibliographystyle{aaai}
\bibliography{reference}

\begin{thebibliography}{}

\bibitem[\protect\citeauthoryear{Becke}{2007}]{becke2007quantum}
Becke, A.
\newblock 2007.
\newblock {\em The quantum theory of atoms in molecules: from solid state to
  DNA and drug design}.
\newblock John Wiley \& Sons.

\bibitem[\protect\citeauthoryear{Behler}{2014}]{behler2014representing}
Behler, J.
\newblock 2014.
\newblock Representing potential energy surfaces by high-dimensional neural
  network potentials.
\newblock {\em Journal of Physics: Condensed Matter} 26(18):183001.

\bibitem[\protect\citeauthoryear{Boomsma and
  Frellsen}{2017}]{boomsma2017spherical}
Boomsma, W., and Frellsen, J.
\newblock 2017.
\newblock Spherical convolutions and their application in molecular modelling.
\newblock In {\em NIPS},  3436--3446.

\bibitem[\protect\citeauthoryear{Bronstein \bgroup et al\mbox.\egroup
  }{2017}]{bronstein2017geometric}
Bronstein, M.~M.; Bruna, J.; LeCun, Y.; Szlam, A.; and Vandergheynst, P.
\newblock 2017.
\newblock Geometric deep learning: going beyond euclidean data.
\newblock {\em IEEE Signal Processing Magazine} 34(4):18--42.

\bibitem[\protect\citeauthoryear{Broomhead and
  Lowe}{1988}]{broomhead1988radial}
Broomhead, D.~S., and Lowe, D.
\newblock 1988.
\newblock Radial basis functions, multi-variable functional interpolation and
  adaptive networks.
\newblock Technical report, Royal Signals and Radar Establishment Malvern
  (United Kingdom).

\bibitem[\protect\citeauthoryear{Collobert and
  Weston}{2008}]{collobert2008unified}
Collobert, R., and Weston, J.
\newblock 2008.
\newblock A unified architecture for natural language processing: Deep neural
  networks with multitask learning.
\newblock In {\em Proceedings of the 25th international conference on Machine
  learning},  160--167.
\newblock ACM.

\bibitem[\protect\citeauthoryear{Cubuk \bgroup et al\mbox.\egroup
  }{2017}]{cubuk2017representations}
Cubuk, E.~D.; Malone, B.~D.; Onat, B.; Waterland, A.; and Kaxiras, E.
\newblock 2017.
\newblock Representations in neural network based empirical potentials.
\newblock {\em The Journal of chemical physics} 147(2):024104.

\bibitem[\protect\citeauthoryear{Faber \bgroup et al\mbox.\egroup
  }{2017}]{faber2017prediction}
Faber, F.~A.; Hutchison, L.; Huang, B.; Gilmer, J.; Schoenholz, S.~S.; Dahl,
  G.~E.; Vinyals, O.; Kearnes, S.; Riley, P.~F.; and von Lilienfeld, O.~A.
\newblock 2017.
\newblock Prediction errors of molecular machine learning models lower than
  hybrid dft error.
\newblock {\em Journal of chemical theory and computation} 13(11):5255--5264.

\bibitem[\protect\citeauthoryear{Gilmer \bgroup et al\mbox.\egroup
  }{2017}]{Gilmer2017NeuralMP}
Gilmer, J.; Schoenholz, S.~S.; Riley, P.~F.; Vinyals, O.; and Dahl, G.~E.
\newblock 2017.
\newblock Neural message passing for quantum chemistry.
\newblock In {\em ICML}.

\bibitem[\protect\citeauthoryear{Goh \bgroup et al\mbox.\egroup
  }{2017}]{goh2017chemnet}
Goh, G.~B.; Siegel, C.; Vishnu, A.; and Hodas, N.~O.
\newblock 2017.
\newblock Chemnet: A transferable and generalizable deep neural network for
  small-molecule property prediction.
\newblock {\em arXiv preprint arXiv:1712.02734}.

\bibitem[\protect\citeauthoryear{G{\'o}mez-Bombarelli \bgroup et
  al\mbox.\egroup }{2018}]{GmezBombarelli2018AutomaticCD}
G{\'o}mez-Bombarelli, R.; Duvenaud, D.~K.; Hern{\'a}ndez-Lobato, J.~M.;
  Aguilera-Iparraguirre, J.; Hirzel, T.~D.; Adams, R.~P.; and Aspuru-Guzik, A.
\newblock 2018.
\newblock Automatic chemical design using a data-driven continuous
  representation of molecules.
\newblock In {\em ACS central science}.

\bibitem[\protect\citeauthoryear{Hansen \bgroup et al\mbox.\egroup
  }{2015}]{hansen2015machine}
Hansen, K.; Biegler, F.; Ramakrishnan, R.; Pronobis, W.; Von~Lilienfeld, O.~A.;
  Müller, K.-R.; and Tkatchenko, A.
\newblock 2015.
\newblock Machine learning predictions of molecular properties: Accurate
  many-body potentials and nonlocality in chemical space.
\newblock {\em The journal of physical chemistry letters} 6(12):2326--2331.

\bibitem[\protect\citeauthoryear{He \bgroup et al\mbox.\egroup
  }{2016}]{He2016DeepRL}
He, K.; Zhang, X.; Ren, S.; and Sun, J.
\newblock 2016.
\newblock Deep residual learning for image recognition.
\newblock {\em CVPR}  770--778.

\bibitem[\protect\citeauthoryear{Hohenberg and
  Kohn}{1964}]{hohenberg1964inhomogeneous}
Hohenberg, P., and Kohn, W.
\newblock 1964.
\newblock Inhomogeneous electron gas.
\newblock {\em Physical review} 136(3B):B864.

\bibitem[\protect\citeauthoryear{Huang and von
  Lilienfeld}{2016}]{huang2016communication}
Huang, B., and von Lilienfeld, O.~A.
\newblock 2016.
\newblock Communication: Understanding molecular representations in machine
  learning: The role of uniqueness and target similarity.

\bibitem[\protect\citeauthoryear{Huang \bgroup et al\mbox.\egroup
  }{2017}]{huang2017question}
Huang, Z.; Liu, Q.; Chen, E.; Zhao, H.; Gao, M.; Wei, S.; Su, Y.; and Hu, G.
\newblock 2017.
\newblock Question difficulty prediction for reading problems in standard
  tests.
\newblock In {\em AAAI},  1352--1359.

\bibitem[\protect\citeauthoryear{Karpathy \bgroup et al\mbox.\egroup
  }{2014}]{Karpathy2014LargeScaleVC}
Karpathy, A.; Toderici, G.; Shetty, S.; Leung, T.; Sukthankar, R.; and Fei-Fei,
  L.
\newblock 2014.
\newblock Large-scale video classification with convolutional neural networks.
\newblock {\em CVPR}  1725--1732.

\bibitem[\protect\citeauthoryear{Kearnes \bgroup et al\mbox.\egroup
  }{2016}]{kearnes2016molecular}
Kearnes, S.; McCloskey, K.; Berndl, M.; Pande, V.; and Riley, P.
\newblock 2016.
\newblock Molecular graph convolutions: moving beyond fingerprints.
\newblock {\em Journal of computer-aided molecular design} 30(8):595--608.

\bibitem[\protect\citeauthoryear{Kingma and Ba}{2014}]{Kingma2014Adam}
Kingma, D.~P., and Ba, J.
\newblock 2014.
\newblock Adam: A method for stochastic optimization.
\newblock {\em Computer Science}.

\bibitem[\protect\citeauthoryear{Kohn and Sham}{1965}]{kohn1965self}
Kohn, W., and Sham, L.~J.
\newblock 1965.
\newblock Self-consistent equations including exchange and correlation effects.
\newblock {\em Physical review} 140(4A):A1133.

\bibitem[\protect\citeauthoryear{Kollman}{1985}]{kollman1985theory}
Kollman, P.
\newblock 1985.
\newblock Theory of complex molecular interactions: computer graphics, distance
  geometry, molecular mechanics, and quantum mechanics.
\newblock {\em Accounts of Chemical Research} 18(4):105--111.

\bibitem[\protect\citeauthoryear{Lawless and
  Chandrasekara}{2002}]{lawless2002information}
Lawless, W., and Chandrasekara, R.
\newblock 2002.
\newblock Information density functional theory: A quantum approach to intent.
\newblock In {\em Proceedings AAAI Fall Conference}.

\bibitem[\protect\citeauthoryear{LeCun, Bengio, and
  others}{1995}]{lecun1995convolutional}
LeCun, Y.; Bengio, Y.; et~al.
\newblock 1995.
\newblock Convolutional networks for images, speech, and time series.
\newblock {\em The handbook of brain theory and neural networks} 3361(10):1995.

\bibitem[\protect\citeauthoryear{Li, Han, and Wu}{2018}]{Li2018DeeperII}
Li, Q.; Han, Z.; and Wu, X.-M.
\newblock 2018.
\newblock Deeper insights into graph convolutional networks for semi-supervised
  learning.
\newblock {\em CoRR} abs/1801.07606.

\bibitem[\protect\citeauthoryear{Liu \bgroup et al\mbox.\egroup
  }{2018}]{liu2018finding}
Liu, Q.; Huang, Z.; Huang, Z.; Liu, C.; Chen, E.; Su, Y.; and Hu, G.
\newblock 2018.
\newblock Finding similar exercises in online education systems.
\newblock In {\em Proceedings of the 24th ACM SIGKDD International Conference
  on Knowledge Discovery \& Data Mining},  1821--1830.
\newblock ACM.

\bibitem[\protect\citeauthoryear{McDonagh \bgroup et al\mbox.\egroup
  }{2017}]{mcdonagh2017machine}
McDonagh, J.~L.; Silva, A.~F.; Vincent, M.~A.; and Popelier, P.~L.
\newblock 2017.
\newblock Machine learning of dynamic electron correlation energies from
  topological atoms.
\newblock {\em Journal of chemical theory and computation} 14(1):216--224.

\bibitem[\protect\citeauthoryear{Montavon \bgroup et al\mbox.\egroup
  }{2012}]{montavon2012learning}
Montavon, G.; Hansen, K.; Fazli, S.; Rupp, M.; Biegler, F.; Ziehe, A.;
  Tkatchenko, A.; Lilienfeld, A.~V.; and M{\"u}ller, K.-R.
\newblock 2012.
\newblock Learning invariant representations of molecules for atomization
  energy prediction.
\newblock In {\em Advances in Neural Information Processing Systems},
  440--448.

\bibitem[\protect\citeauthoryear{Oglic, Garnett, and
  G{\"a}rtner}{2017}]{oglic2017active}
Oglic, D.; Garnett, R.; and G{\"a}rtner, T.
\newblock 2017.
\newblock Active search in intensionally specified structured spaces.
\newblock In {\em AAAI},  2443--2449.

\bibitem[\protect\citeauthoryear{Ramakrishnan and von
  Lilienfeld}{2015}]{ramakrishnan2015many}
Ramakrishnan, R., and von Lilienfeld, O.~A.
\newblock 2015.
\newblock Many molecular properties from one kernel in chemical space.
\newblock {\em CHIMIA International Journal for Chemistry} 69(4):182--186.

\bibitem[\protect\citeauthoryear{Ramakrishnan \bgroup et al\mbox.\egroup
  }{2014}]{ramakrishnan2014quantum}
Ramakrishnan, R.; Dral, P.~O.; Rupp, M.; and von Lilienfeld, O.~A.
\newblock 2014.
\newblock Quantum chemistry structures and properties of 134 kilo molecules.
\newblock {\em Scientific Data} 1.

\bibitem[\protect\citeauthoryear{Ruddigkeit \bgroup et al\mbox.\egroup
  }{2012}]{Ruddigkeit2012EnumerationO1}
Ruddigkeit, L.; van Deursen, R.; Blum, L.~C.; and Reymond, J.-L.
\newblock 2012.
\newblock Enumeration of 166 billion organic small molecules in the chemical
  universe database gdb-17.
\newblock {\em Journal of chemical information and modeling} 52 11:2864--75.

\bibitem[\protect\citeauthoryear{Sch{\"u}tt \bgroup et al\mbox.\egroup
  }{2017a}]{schutt2017schnet}
Sch{\"u}tt, K.; Kindermans, P.-J.; Felix, H. E.~S.; Chmiela, S.; Tkatchenko,
  A.; and M{\"u}ller, K.-R.
\newblock 2017a.
\newblock Schnet: A continuous-filter convolutional neural network for modeling
  quantum interactions.
\newblock In {\em NIPS},  992--1002.

\bibitem[\protect\citeauthoryear{Sch{\"u}tt \bgroup et al\mbox.\egroup
  }{2017b}]{schutt2017quantum}
Sch{\"u}tt, K.~T.; Arbabzadah, F.; Chmiela, S.; M{\"u}ller, K.~R.; and
  Tkatchenko, A.
\newblock 2017b.
\newblock Quantum-chemical insights from deep tensor neural networks.
\newblock {\em Nature communications} 8:13890.

\bibitem[\protect\citeauthoryear{Shang \bgroup et al\mbox.\egroup
  }{2018}]{shang2018edge}
Shang, C.; Liu, Q.; Chen, K.-S.; Sun, J.; Lu, J.; Yi, J.; and Bi, J.
\newblock 2018.
\newblock Edge attention-based multi-relational graph convolutional networks.
\newblock {\em arXiv preprint arXiv:1802.04944}.

\bibitem[\protect\citeauthoryear{Thouless}{2014}]{thouless2014quantum}
Thouless, D.~J.
\newblock 2014.
\newblock {\em The quantum mechanics of many-body systems}.
\newblock Courier Corporation.

\bibitem[\protect\citeauthoryear{Wang and Hou}{2011}]{wang2011application}
Wang, J., and Hou, T.
\newblock 2011.
\newblock Application of molecular dynamics simulations in molecular property
  prediction ii: diffusion coefficient.
\newblock {\em Journal of computational chemistry} 32(16):3505--3519.

\bibitem[\protect\citeauthoryear{Yanai, Tew, and Handy}{2004}]{yanai2004new}
Yanai, T.; Tew, D.~P.; and Handy, N.~C.
\newblock 2004.
\newblock A new hybrid exchange--correlation functional using the
  coulomb-attenuating method (cam-b3lyp).
\newblock {\em Chemical Physics Letters} 393(1-3):51--57.

\bibitem[\protect\citeauthoryear{Zhu \bgroup et al\mbox.\egroup
  }{2016}]{zhu2016co}
Zhu, W.; Lan, C.; Xing, J.; Zeng, W.; Li, Y.; Shen, L.; Xie, X.; et~al.
\newblock 2016.
\newblock Co-occurrence feature learning for skeleton based action recognition
  using regularized deep lstm networks.
\newblock In {\em AAAI}, volume~2, ~6.

\bibitem[\protect\citeauthoryear{Zhu \bgroup et al\mbox.\egroup
  }{2018}]{zhu2018xiaoice}
Zhu, H.; Liu, Q.; Yuan, N.~J.; Qin, C.; Li, J.; Zhang, K.; Zhou, G.; Wei, F.;
  Xu, Y.; and Chen, E.
\newblock 2018.
\newblock Xiaoice band: A melody and arrangement generation framework for pop
  music.
\newblock In {\em Proceedings of the 24th ACM SIGKDD International Conference
  on Knowledge Discovery \& Data Mining},  2837--2846.
\newblock ACM.

\bibitem[\protect\citeauthoryear{Zou and Hastie}{2005}]{zou2005regularization}
Zou, H., and Hastie, T.
\newblock 2005.
\newblock Regularization and variable selection via the elastic net.
\newblock {\em Journal of the Royal Statistical Society: Series B (Statistical
  Methodology)} 67(2):301--320.

\end{thebibliography}

\end{document}